\documentclass[10pt,a4paper,reqno]{amsart}
\usepackage{acronym}
\usepackage{amsfonts}
\usepackage{amsmath}
\usepackage{amssymb}
\usepackage{amsthm}
\usepackage{bm}
\usepackage{braket}
\usepackage{cite}
\usepackage{color}
\usepackage{geometry}
\usepackage{graphicx}
\usepackage{hyperref}
\usepackage{mathrsfs}
\usepackage{mathtools}
\usepackage{setspace}
\usepackage{stmaryrd}
\usepackage{subcaption}
\usepackage{tikz}

\usepackage{epsfig}
\usepackage{setspace}
\usepackage{booktabs}
\usepackage{threeparttable}
\usepackage{diagbox}

\newgeometry{top=2cm,bottom=2cm,outer=1.5cm,inner=1.5cm}
\newcommand{\bse}{\begin{subequations}}
\newcommand{\ese}{\end{subequations}}

\numberwithin{equation}{section}

\title[Soliton, Breather and Rogue Wave Solutions for Solving the Nonlinear Schr\"{o}dinger Equation Using a Deep Learning Method with Physical Constraints]{Soliton, Breather and Rogue Wave Solutions for Solving the Nonlinear Schr\"{o}dinger Equation Using a Deep Learning Method with Physical Constraints}
\author{Juncai Pu}
\address[JP]{School of Mathematical Sciences, Shanghai Key Laboratory of Pure Mathematics and Mathematical Practice, and Shanghai Key Laboratory of Trustworthy Computing \\
East China Normal University \\ Shanghai 200241 \\ People's Republic of China}
\author{Jun Li}
\address[JL]{Shanghai Key Laboratory of Trustworthy Computing \\ East China Normal University \\ Shanghai 200062\\
People's Republic of China}
\author{Yong Chen$^*$}
\address[YC]{School of Mathematical Sciences, Shanghai Key Laboratory of Pure Mathematics and Mathematical Practice, and Shanghai Key Laboratory of Trustworthy Computing \\
East China Normal University \\ Shanghai 200241 \\ People's Republic of China}
\address[YC]{College of Mathematics and Systems Science \\ Shandong University of Science and Technology \\ Qingdao 266590 \\ People's Republic of China}
\address[YC]{Department of Physics \\ Zhejiang Normal University \\ Jinhua 321004 \\ People's Republic of China}
\email{ychen@sei.ecnu.edu.cn}

\begin{document}

\begin{abstract}
The nonlinear Schr\"odinger equation is a classical integrable equation which contains plenty of significant properties and occurs in many physical areas. However, due to the difficulty of solving this equation, in particular in high dimensions, lots of methods are proposed to effectively obtain different kinds of solutions, such as neural networks, among others. Recently, a method where some underlying physical laws are embeded into a conventional neural network is proposed to uncover the equation's dynamical behaviors from spatiotemporal data directly. Compared with traditional neural networks, this method can obtain remarkably accurate solution with extraordinarily less data. Meanwhile, this method also provides a better physical explanation and generalization. In this paper, based on the above method, we present an improved deep learning method to recover the soliton solutions, breather solution and rogue wave solutions of the nonlinear Schr\"odinger equation. In particular, the dynamical behaviors and error analysis about the one-order and two-order rogue waves of nonlinear integrable equations are revealed by the deep neural network with physical constraints for the first time. Moreover, the effects of different numbers of initial points sampled, residual collocation points sampled, network layers, neurons per hidden layer on the one-order rogue wave dynamics of this equation have been considered with the help of the control variable way under the same initial and boundary conditions. Numerical experiments show that the dynamical behaviors of soliton solutions, breather solution and rogue wave solutions of the integrable nonlinear Schr\"odinger equation can be well reconstructed by utilizing this physically-constrained deep learning method.
\end{abstract}

\maketitle

\section{Introduction}

In recent decades, more and more attention have been paid to the nonlinear problems in fluid mechanics, condensed matter physics, optical fiber communication, plasma physics and even biology \cite{Draper1964, Peregrine1983, Zabusky1965, Parkins1998}. After establishing nonlinear partial
differential equations to describe these nonlinear phenomena and then analyzing the analytical and numerical solutions of these nonlinear models, the essence of these nonlinear phenomena can be understood \cite{Ablowitz1992}. Therefore, the research
of these nonlinear problems is essentially transformed into the study of nonlinear partial differential equations which describe these physical phenomena. Due to many basic properties of linear differential equations are not applicable to nonlinear differential equations, these nonlinear differential equations which the famous nonlinear Schr\"{o}dinger equation belongs to are more difficult to solve compared with the linear differential equations. It is well known that the Schr\"{o}dinger equation can be used to describe the quantum behavior of microscopic particles in quantum mechanics \cite{Schr1926}. Furthermore, various solutions of this equation can describe the nonlinear phenomena in other physical fields, such as optical fiber, plasma, Bose-Einstein condensates, fluid mechanics and Heisenberg ferromagnet \cite{Guo2012,Solli2007,Chabchoub2011,Qiao1994,Akhmediev2009,Ohta2012,Hasegawa1973,Kavitha2003,Qiao1993}.

With the explosive growth of available data and computing resources, deep neural networks, i.e., deep learning \cite{LeCun2015}, are applied in many areas including image recognition, video surveillance, natural language processing, medical diagnostics, bioinformatics, financial data analysis and so on \cite{Mitchell1997,Bishop2006,Alipanahi2015,Duda2000,Lake2015,Krizhevsky2017}. In scientific computing, especially, the neural network method \cite{McCulloch1943,Rosenblatt1958,Bryson1975} provides an ideal representation for the solution of differential equations \cite{Lagaris1998} due to its universal approximation properties \cite{Hornik1989}. Recently, a physically constrained deep learning method called physics-informed neural network (PINN) \cite{Raissi2019} and its improvement \cite{Jagtap2020} has been proposed which is particularly suitable for solving differential equations and corresponding inverse problems. It is found that the PINN architecture can obtain remarkably accurate solution with extraordinarily less data. Meanwhile, this method also provides a better physical explanation for predicted solutions because of the underlying physical constraints which is usually described explicitly by the differential equations. In this paper, the computationally efficient physics-informed data-driven algorithm for inferring solutions to more general nonlinear partial differential equations, such as the integrable nonlinear Schr\"odinger equation, is studied.

As is known to all, the study of exact solutions for integrable equations which are used to describe complex physical phenomena in the real world have been paid more and more attention in plasma physics, optical fiber, fluid dynamics and others fields \cite{Lax1968,Hasegawa1973,Iwao1997,YuS1998,Osman2019}. The Hirota bilinear method, the symmetry reduction method, the Darboux transformation, the B$\mathrm{\ddot{a}}$cklund transformations, the inverse scattering method and the function expansion method are powerful means to solve nonlinear integrable equations, and many other methods are based on them \cite{Hirota2004,Geng1999,Matveev1991,Olver1993,Zakharov1984,Ablowitz1992,Pu2020}. Although the computational cost of some direct numerical solutions of integrable equations is very high, with the revival of neural networks, the development of more effective deep learning algorithms to obtain data-driven solutions of nonlinear integrable equations has aroused great interest \cite{Li2020,Bongard2007,Raissi2017,Lij2020,LiJ2020}. Li and Chen constructed abundant numerical solutions of second-order and third-order nonlinear integrable equations with different initial and boundary conditions by deep learning method based on the PINN model \cite{Li2020,Lij2020,LiJ2020}. Previous works mainly focused on some simple solutions (e.g., N-soliton solutions, kink solutions) of given system or integrable equation. Relatively, the research results of machine learning for constructing rogue waves are rare. In Ref. \cite{Marcucci2020}, the bias function including two backward shock waves and soliton generation and the generation of rogue waves are studied by using a single wave-layer feed forward neural network.  As far as we know, the soliton solutions, breather solution and rogue wave solutions \cite{Solli2007,Chabchoub2011} of the integrable nonlinear Schr\"odinger equation have not been given out by the deep learning method based on PINN. Therefore, we introduce the deep learning method with underlying physical constraints to construct the soliton solutions, breathing solution and rogue wave solutions of integrable nonlinear Schr\"odinger equation in this work.

This paper is organized as follows. In section 2, we introduce the physically constrained deep learning method and
briefly present some problem setups. In Section 3, the one-soliton solution and two-soliton solution of the nonlinear Schr\"odinger equation is obtained by this approach, and the breather solution are derived compared with the two-soliton solution. Section 4 provides rogue wave solutions which contain one-order rogue wave and two-order rogue wave for the nonlinear Schr\"odinger equation, and the relative $\mathbb{L}_2$ errors of simulating the one-order rogue wave of nonlinear Schr\"odinger equation with different numbers of initial points sampled, residual collocation points sampled, network layers and neurons per hidden layer are also given out in detail. Conclusion is given in last section.

\section{Method}

In this paper, we consider (1+1)-dimensional nonlinear Schr\"odinger equation as follows
\begin{equation}\label{E1}
i q_t+\alpha q_{xx}+\beta |q|^2 q=0,
\end{equation}
where $\alpha, \beta$ are arbitrary parameters, $i = \sqrt{-1}$ and $q$ are complex-valued solutions with respect to $x, t$. Based on the theory of integrable systems and PINN, we establish a physically-constrained deep learning method to approximate the potential solution $|q(x,t)|$ of this integrable equation. Here, the underlying laws of physics are described explicitly by this equation and embedded into the architecture with the help of automatic differentiation \cite{Baydin2018}. The physical constraints regarded as a kind of regularization are introduced into the network via this mechanism which enables us to understand the network architecture and the predicted solutions better. In addition, due to the physical constraints, the network is trained just with few data.

Specifically, the complex value solution $q(x,t)$ is formulated as $q=u+i v$, where $u(x,t)$ and $v(x,t)$ are real-valued functions of $x,t$, and real part and imaginary part of $q(x,t)$, respectively. Then, Eq. \eqref{E1} can be converted into
\begin{equation}\label{E2}
u_t+\alpha v_{xx} + \beta (u^2+v^2)v=0,
\end{equation}
\begin{equation}\label{E3}
v_t-\alpha u_{xx} - \beta (u^2+v^2)u=0.
\end{equation}

Accordingly, We define the residuals $f_u(x, t)$ and $f_v(x, t)$ respectively
\begin{equation}\label{E4}
f_u := u_t+\alpha v_{xx} + \beta (u^2+v^2)v,
\end{equation}
\begin{equation}\label{E5}
f_v := v_t-\alpha u_{xx} - \beta (u^2+v^2)u,
\end{equation}
and the solution $q(x,t)$ is trained to satisfy the networks \eqref{E4} and \eqref{E5} which are embedded into the mean-squared objective function (also called loss function)
\begin{equation}\label{E6}
Loss=Loss_u+Loss_v+Loss_{f_u}+Loss_{f_v},
\end{equation}
where
\begin{equation}\label{E7}
Loss_u=\frac{1}{N_q}\sum^{N_q}_{i=1}|u(x_u^i,t_u^i)-u^i|^2,
\end{equation}
\begin{equation}\label{E8}
Loss_v=\frac{1}{N_q}\sum^{N_q}_{i=1}|v(x_v^i,t_v^i)-v^i|^2,
\end{equation}
\begin{equation}\label{E9}
Loss_{f_u}=\frac{1}{N_f}\sum^{N_f}_{j=1}|f_u(x_{f_u}^j,t_{f_u}^j)|^2,
\end{equation}
and
\begin{equation}\label{E10}
Loss_{f_v}=\frac{1}{N_f}\sum^{N_f}_{j=1}|f_v(x_{f_v}^j,t_{f_v}^j)|^2.
\end{equation}
Here the initial and boundary value data about $q(x,t)$ are denoted by $\{x^i_u,t^i_u,u^i\}^{N_q}_{i=1}$ and $\{x^i_v,t^i_v,v^i\}^{N_q}_{i=1}$. Similarly, the collocation points for $f_u(x,t)$ and $f_v(x,t)$ are specified as $\{x_{f_u}^j,t_{f_u}^j\}^{N_{f}}_{j=1}$ and $\{x_{f_v}^j,t_{f_v}^j\}^{N_{f}}_{j=1}$ which are sampled using the classical latin hypercube sampling (LHS) technique \cite{Stein1987}. The Loss function \eqref{E6} corresponds to the initial-boundary data and the residuals imposed by Eq. \eqref{E4} and \eqref{E5} at a finite set of collocation points sampled. Specifically, the first and second terms on the right hand side of Eq. \eqref{E6} attempt to fit the solution data, and the third and fourth terms learn to satisfy the residuals $f_u$ and $f_v$. The convergence of the loss function has been analyzed in previous works \cite{Choromanska2015}.

In this paper, we optimize all loss functions simply using the L-BFGS algorithm which is a full-batch gradient-based optimization algorithm based on a quasi-Newton method \cite{Liu1989}. In addition, we use relatively simple multilayer perceptrons (MLPs) with the Xavier initialization and the hyperbolic tangent ($\tanh$) activation function \cite{Li2020}. All codes in this article are based on Python 3.7 and Tensorflow 1.15, and all numerical examples reported here are run on a DELL Precision 7920 Tower computer with 2.10 GHz 8-core Xeon Silver 4110 processor and 64 GB memory.

\section{Soliton solutions and breather solution of the nonlinear Schr\"odinger equation}

The (1+1)-dimensional focusing nonlinear Schr\"odinger equation is a classical integrable field equation for describing quantum mechanical systems, nonlinear wave propagation in optical fibers or waveguides, Bose-Einstein condensates and plasma waves. In optics, the nonlinear term is generated by the intensity dependent index of a given material. Similarly, the nonlinear term for Bose-Einstein condensates is the result of the mean-field interactions about the interacting N-body system. We consider the focusing nonlinear Schr\"odinger equation along with Dirichlet boundary conditions given by
\begin{equation}\label{E11}
\begin{split}
\begin{cases}
iq_t+q_{xx}+2|q|^2q=0,x\in[x_0,x_1],t\in[t_0,t_1],\\
q(x,t_0)=q_0(x),\\
q(x_0,t)=q(x_1,t),\\
\end{cases}
\end{split}
\end{equation}
where $q_0(x)$ is an arbitrary complex-valued function of space variable $x$, $x_0,x_1$ represent the lower and upper boundaries of $x$ respectively, and $t_0,t_1$ represent the initial and terminal time instants of $t$ respectively. In addition, this equation corresponds to Eq. \eqref{E1} with $\alpha=1$ and $\beta=2$. Eq. \eqref{E11} is often used to describe the evolution of weakly nonlinear dispersive wave modulation. In view of the characteristic of its solution, it is called "self focusing" nonlinear Schr\"odinger equation. For water wave modulation, there is usually coupling between modulation and wave induced current, so in some cases, water wave modulation can also be described by the nonlinear Schr\"odinger equation \cite{Peregrine1983}. The N-soliton solutions and breather solution of the above equation have been obtained by many different methods \cite{Hirota2004,Matveev1991,Yang2010}. Here, we simulate the soliton solutions and breather solution using the physically constrained deep learning method, and compare them with the known exact solutions, so as to prove the effectiveness of solving the numerical solutions $q(x,t)$ by neural networks. Specifically, the N-soliton solution of nonlinear Schr\"odinger equation have been derived by the Riemann Hilbert method \cite{Yang2010}, and the N-soliton solution is formed as
\begin{equation}\label{E12}
q(x,t)=-2i\frac{\mathrm{det}R}{\mathrm{det}M},
\end{equation}
where $M$ is a matrix of $N\times N$,
\begin{equation}\label{E13}
M=\begin{bmatrix}  M_{11} & M_{12} & \cdots &M_{1N} \\ M_{21} & M_{22} & \cdots &M_{2N} \\ \vdots & \vdots &\vdots&\vdots \\ M_{N1} & M_{N2} & \cdots & M_{NN} \\\end{bmatrix},
\end{equation}
with
\begin{equation}\label{E14}
M_{jk}=\frac{e^{-(\theta_k+\theta^*_j)}+c^*_jc_ke^{\theta_k+\theta^*_j}}{\zeta^*_j-\zeta_k},\quad j,k=1\cdots N,
\end{equation}
and $R$ is a matrix of $(N+1)\times(N+1)$,
\begin{equation}\label{E15}
R=\begin{bmatrix} 0 &c_1e^{\theta_1} & \cdots & c_Ne^{\theta_N} \\   e^{-\theta^*_1}& M_{11} &\cdots &M_{1N} \\ \vdots & \vdots &\vdots&\vdots \\ e^{-\theta^*_N} & M_{N1} & \cdots & M_{NN} \\\end{bmatrix},
\end{equation}
with $\theta_k=-i\zeta_kx-2i\zeta^2_kt (k=1\cdots N)$, $\zeta_k$ and $c_i (i=1\cdots N)$ are complex value constants. After taking the positive integer $N$, one can obtain the corresponding N-soliton solutions and breather solution of the nonlinear Schr\"odinger Eq. \eqref{E11}.
\subsection{One-soliton solution}
\quad

In this subsection, we numerically construct one-soliton solution of Eq. \eqref{E11} based on the neural network structure with 9 hidden layers and 40 neurons per hidden layer. When $N=1$, we have $R=\left(\begin{array}{cc}0 & c_1e^{\theta_1} \\ e^{-\theta^*_1} & M_{11} \\ \end{array}\right), M_{11}=\frac{e^{-(\theta_1+\theta^*_1)}+c^*_1c_1e^{\theta_1+\theta^*_1}}{\zeta_1^*-\zeta_1}, \mathrm{det}R=-c_1e^{\theta_1-\theta^*_1}, \mathrm{det}M=M_{11}$, so the solution \eqref{E12} is
\begin{equation}\label{E16}
q(x,t)=-2i\frac{\mathrm{det}R}{\mathrm{det}M}=2i(\zeta_1^*-\zeta_1)\frac{c_1e^{\theta_1-\theta^*_1}}{e^{-(\theta_1+\theta^*_1)}+|c_1|^2e^{\theta_1+\theta^*_1}}.
\end{equation}

Letting $$\zeta_1=\xi+i\eta,\quad c_1=e^{-2\eta x_0+i\sigma_0},$$
where $\xi, \eta$ are the real and imaginary parts of $\zeta_1$ respectively, and $x_0, \sigma_0$ are real parameters. Then the above one-soliton solution \eqref{E16} can be reduced to
\begin{equation}\label{E17}
q(x,t)=2\eta\mathrm{sech}[2\eta(x+4\xi t-x_0)]e^{[-2i\xi x-4i(\xi^2-\eta^2)t+i\sigma_0]}.
\end{equation}

One can obtain the exact one-soliton solution of the nonlinear Schr\"odinger Eq. \eqref{E11} after taking $\eta=1,\xi=1,x_0=0,\sigma_0=1$ into \eqref{E17} as follows
\begin{equation}\label{E18}
q(x,t)=2\mathrm{sech}(8t+2x)e^{(-2ix+i)}.
\end{equation}

Then we take $[x_0,x_1]$ and $[t_0,t_1]$ in Eq. \eqref{E11} as $[-5.0,5.0]$ and $[-0.5,0.5]$, respectively. The corresponding initial condition is obtained by substituting a specific initial value into \eqref{E18}
\begin{equation}\label{E19}
q_0(x)=2\mathrm{sech}(2x-4)e^{(-2ix+i)}.
\end{equation}

We employ the traditional finite difference shcemes on even grids in MATLAB to simulate Eq. \eqref{E11} with the initial data \eqref{E19} to acquire the training data. Specifically, dividing space $[-5.0,5.0]$ into 513 points and time $[-0.5,0.5]$ into 401 points, one-soliton solution $q$ is discretized into $401$ snapshots accordingly. We sub-sample a smaller training dataset that contain initial-boundary subsets by randomly extracting $N_q=100$ from original initial-boundary data and $N_f=10000$ collocation points which are generated by LHS \cite{Stein1987}. After giving a dataset of initial and boundary points, the latent one-soliton solution $q(x,t)$ is successfully learned by tuning all learnable parameters of the neural network and regulating the loss function \eqref{E6}. The model achieves a relative $\mathbb{L}_2$ error of 2.566069$\mathrm{e}-$02 in about $726$ seconds, and the number of iterations is 8324.

In Fig. 1, the density diagrams, the figures at different instants of the latent one-soliton solution $q(x,t)$, the error diagram about the difference between exact one-soliton solution and hidden one-soliton solution, and the loss curve figure are plotted respectively. The (a) of Fig. 1 clearly compares the exact solution with the predicted spatiotemporal solution. Obviously, combining with the (b), we can see that the error between the numerical solution and the exact solution is very small. We particularly present a comparison between the exact solution and the predicted solution at different time instants $t = -0.25, 0, 0.25$  in the bottom panel of (a). It is obvious that as time t increases, the one-soliton solution propagates along the negative direction of the $x$-axis. The three dimensional motion of the predicted solution and the loss curve at different iterations are given out in detail in (c) and (d) of Fig. 1. The results show that the loss curve is very smooth which proves the effectiveness and stability of the integrable deep learning method.
\begin{figure}
\centering
\includegraphics[width=6.5cm,height=5cm]{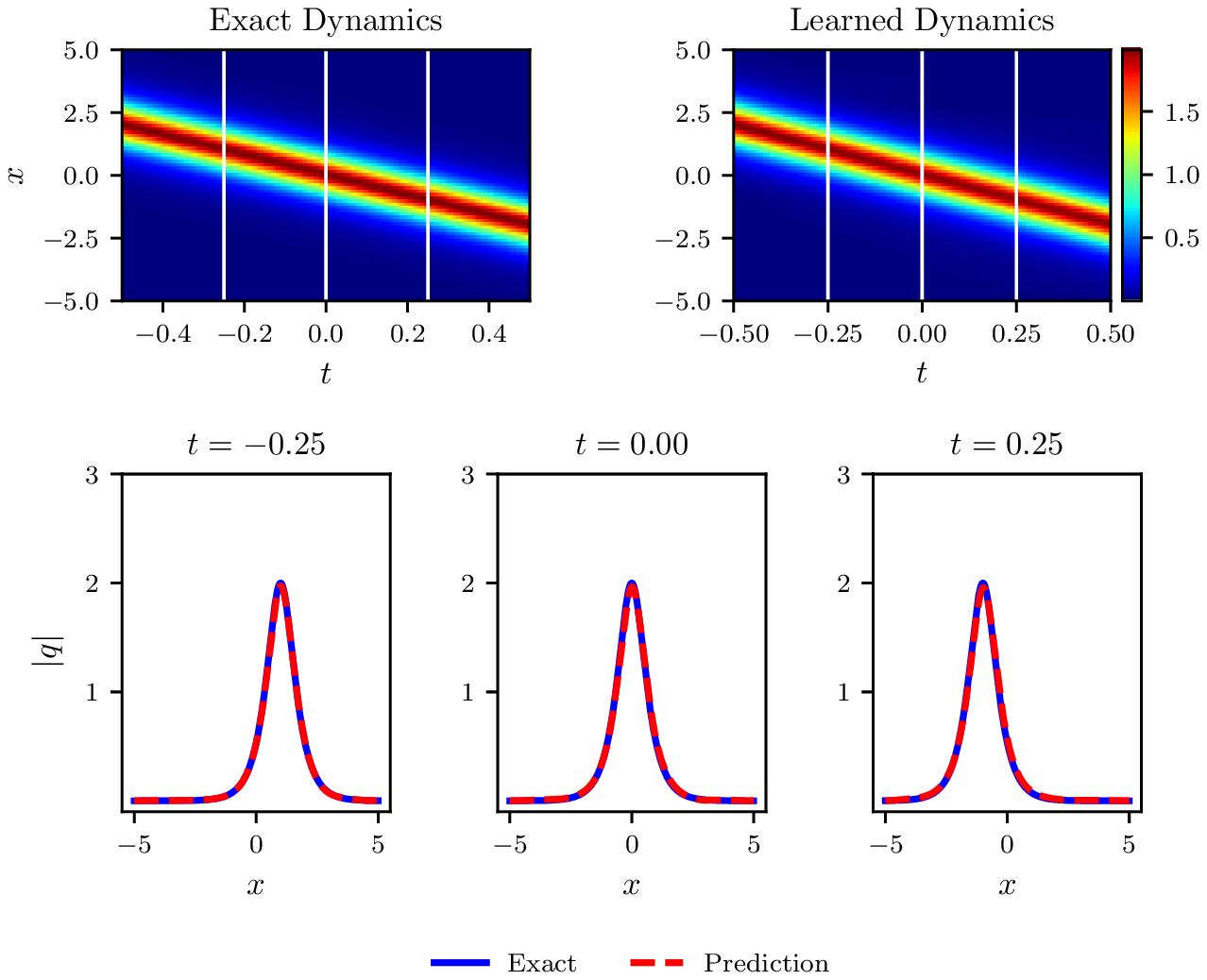}
$a$
\includegraphics[width=6.5cm,height=5cm]{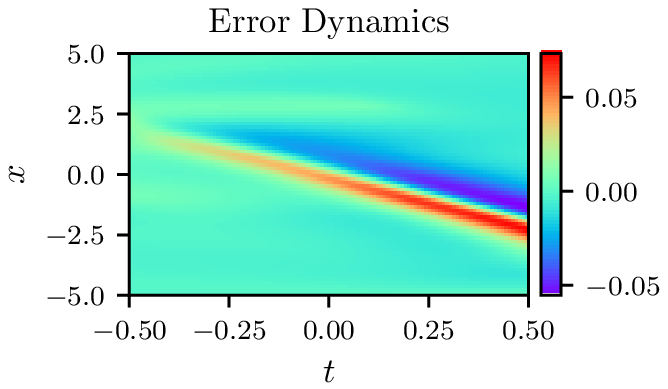}
$b$\\
\includegraphics[width=6.5cm,height=5cm]{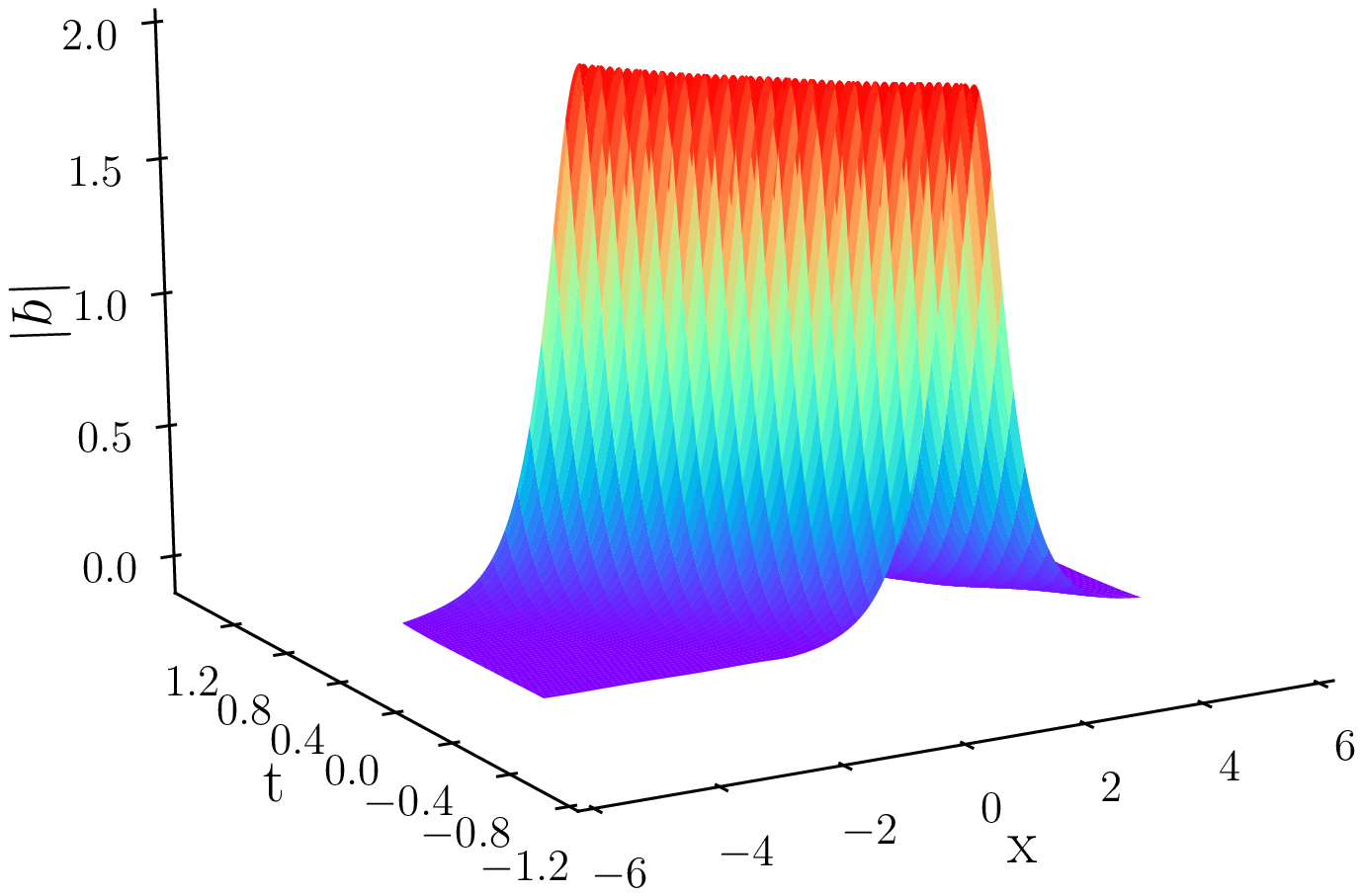}
$c$
\includegraphics[width=6.5cm,height=5cm]{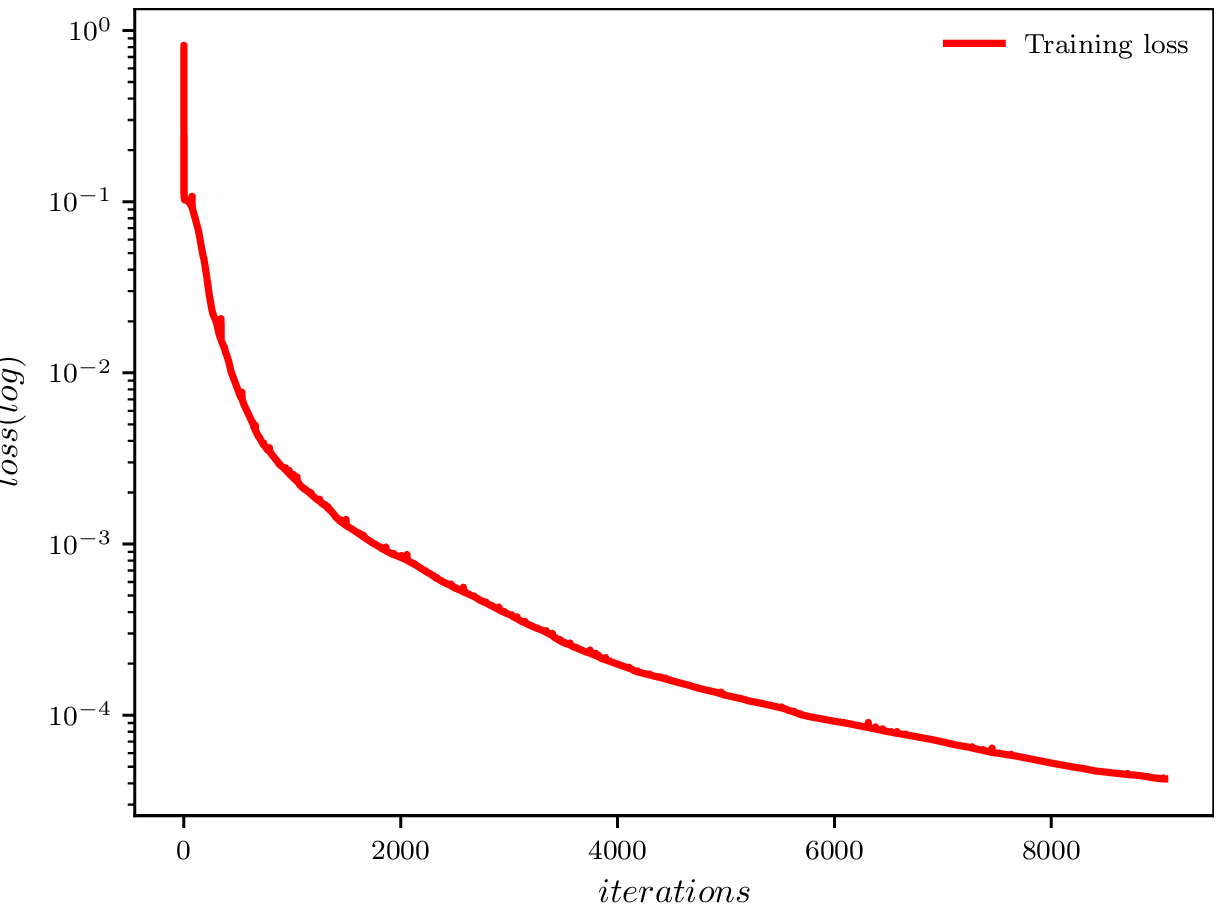}
$d$
\caption{The one-soliton solution $q(x,t)$: (a) The density diagrams and figures at three different instants, respecitvely; (b) The error density diagram; (c) The three-dimensional motion; (d) The loss curve figure.}
\end{figure}

\subsection{Two-soliton solution and breather solution}
\quad

Now, we numerically construct the two-soliton solution and breather solution of Eq. \eqref{E11} based on the neural network architecture with 9 hidden layers and 80 neurons per hidden layer. When $N=2$, the solution \eqref{E12} can also be written out explicitly. We have
$$F=\left(\begin{array}{ccc}0 & c_1e^{\theta_1} & c_2e^{\theta_2} \\ e^{-\theta^*_1} & M_{11} & M_{12} \\  e^{-\theta^*_2} & M_{21} & M_{22} \\ \end{array}\right), M=\left(\begin{array}{cc}M_{11} & M_{12} \\ M_{21} & M_{22} \\ \end{array}\right),$$
where
\begin{equation}\nonumber
\begin{split}
M_{11}=\frac{e^{-(\theta_1+\theta^*_1)}+c^*_1c_1e^{\theta_1+\theta^*_1}}{\zeta_1^*-\zeta_1}, M_{12}=\frac{e^{-(\theta_2+\theta^*_1)}+c^*_1c_2e^{\theta_2+\theta^*_1}}{\zeta_1^*-\zeta_2},\\
M_{21}=\frac{e^{-(\theta_1+\theta^*_2)}+c^*_2c_1e^{\theta_1+\theta^*_2}}{\zeta_2^*-\zeta_1},
M_{22}=\frac{e^{-(\theta_2+\theta^*_2)}+c^*_2c_2e^{\theta_2+\theta^*_2}}{\zeta_2^*-\zeta_2},\\
\end{split}
\end{equation}
with $\theta_1=-i\zeta_1x-2i\zeta^2_1t, \theta^*_1=i\zeta^*_1x+2i\zeta^{*2}_1t, \theta_2=-i\zeta_2x-2i\zeta^2_2t, \theta^*_2=-i\zeta^*_2x-2i\zeta^{*2}_2t$, $\zeta_j (j=1,2)$ are complex value constants, so one can derive the general form of two-soliton solution as follows
\begin{equation}\label{E20}
\begin{split}
q(x,t)=2i\frac{c_2M_{11}e^{\theta_2-\theta^*_2}-c_2M_{12}e^{\theta_2-\theta^*_1}-c_1M_{21}e^{\theta_1-\theta^*_2}+c_1M_{22}e^{\theta_1-\theta^*_1}}{M_{11}M_{22}-M_{12}M_{21}}.
\end{split}
\end{equation}

According to the relationship between the two-soliton solution and the breather solution, we can know that when $\mathrm{Re}(\zeta_1)\neq\mathrm{Re}(\zeta_2)$, the solution $q(x,t)$ is a two-soliton solution, and when $\mathrm{Re}(\zeta_1)=\mathrm{Re}(\zeta_2)$, the solution $q(x,t)$ degenerates into a bound state which is also called the breather solution. Given appropriate parameters
\begin{equation}\label{E21}
\zeta_1=0.1+0.7i,\quad\zeta_2=-0.1+0.4i,\quad c_1=c_2=1,
\end{equation}
we can obtain the exact two-soliton solution from the formulae \eqref{E20}
\begin{equation}\label{E22}
q(x,t)=\frac{-2iA}{B},
\end{equation}
where
\begin{equation}\nonumber
\begin{split}
A=&(0.224+0.158i)e^{i(1.92t+0.32it-0.2x-0.8ix)}-(0.224+0.232i)e^{-i(-0.6t+0.56it-0.2x+1.4ix)}+\\
&(0.224-0.232i)e^{i(0.6t+0.56it+0.2x+1.4ix)}+(0.518i-0.224)e^{-i(-1.92t+0.32it+0.2x-0.8ix)},
\end{split}
\end{equation}
and
\begin{equation}\nonumber
\begin{split}
B=1.25e^{0.88t+0.6x}+0.13e^{-0.24t-2.2x}+1.25e^{-0.88t-0.6x}-1.12e^{i(1.32t-0.4x)}-1.12e^{-i(1.32t-0.4x)}+0.13e^{0.24t+2.2x}.
\end{split}
\end{equation}

On the other hand, given other appropriate parameters
\begin{equation}\label{E23}
\zeta_1=0.7i,\quad\zeta_2=0.4i,\quad c_1=c_2=1,
\end{equation}
one can obtain the exact breather solution
\begin{equation}\label{E24}
q(x,t)=\frac{-2i(-0.246ie^{-i(-0.64t+1.4ix)}+0.462ie^{-i(-1.96t-0.8ix)}-0.264ie^{i(0.64t+1.4ix)}+0.462ie^{-i(-1.96t+0.8ix)})}{1.21e^{-0.6x}+1.21e^{0.6x}-1.12e^{1.32it}+0.09e^{2.2x}-1.12e^{-1.32it}+0.09e^{-2.2x}}.
\end{equation}

Now we take $[x_0,x_1]$ and $[t_0,t_1]$ in Eq. \eqref{E11} as $[-5.0,5.0]$ and $[-3.0,3.0]$, respectively. For instance, we consider the initial condition of the two-soliton solution based on Eq. \eqref{E22}
\begin{equation}\label{E25}
q_0(x)=\frac{-2iA'}{B'},
\end{equation}
where
\begin{equation}\nonumber
\begin{split}
A'=&(0.224+0.158i)e^{i(-5.76-0.96i-0.2x-0.8ix)}-(0.224+0.232i)e^{-i(1.8-1.68i-0.2x+1.4ix)}+\\
&(0.224-0.232i)e^{i(-1.8-1.68i+0.2x+1.4ix)}+(0.518i-0.224)e^{-i(5.76-0.96i+0.2x-0.8ix)},
\end{split}
\end{equation}
and
\begin{equation}\nonumber
\begin{split}
B'=1.25e^{-2.64+0.6x}+0.13e^{0.72-2.2x}+1.25e^{2.64-0.6x}-1.12e^{i(-3.96-0.4x)}-1.12e^{-i(-3.96-0.4x)}+0.13e^{-0.72+2.2x}.
\end{split}
\end{equation}

Similarly, the initial condition of the breather solution is given
\begin{equation}\label{E26}
q_0(x)=\frac{-2i\left(-0.246ie^{-i(1.92+1.4ix)}+0.462ie^{-i(5.88-0.8ix)}-0.264ie^{i(-1.92+1.4ix)}+0.462ie^{-i(5.88+0.8ix)}\right)}{1.21e^{-0.6x}+1.21e^{0.6x}-1.12e^{-3.96i}+0.09e^{2.2x}-1.12e^{3.96i}+0.09e^{-2.2x}}.
\end{equation}

With the same data generation and sampling method in Section 3.1, we numerically simulate the two-soliton solution and the breather solution of the nonlinear Schr\"odinger equation \eqref{E11} using the physically-constrained deep learning method mentioned above. After training the two-soliton solution, the neural network achieves a relative $\mathbb{L}_2$ error of 5.500792$\mathrm{e}-$02 in about $2565$ seconds, and the number of iterations is 17789. However, the network model for learning breather solution achieves a relative $\mathbb{L}_2$ error of 9.689267$\mathrm{e}-$03 in about $1934$ seconds, and the number of iterations is 13488. Apparently, since the breather solution is a special form of the two-soliton solution and accordingly the solution structure is simpler, the training of the breather solution takes remarkably less time, the relative error is obviously smaller, and moreover the result is better than that of the two-soliton solution from Fig. 2 and Fig. 3.

Fig. 2 and Fig. 3 show the density diagrams, the profiles at different instants and error density diagrams of the two-soliton solution and the breather solution, respectively. From the bottom panel of $(a)$ in Fig. 2, we can clearly see that the intersection of two solitary waves with different wave widths and amplitudes produces a peak of a higher amplitude different from the former two solitary waves, which satisfies the law of conservation of energy. We reveal the profiles of the three moments at $t=-1.50,0,1.50$, respectively, and find that the amplitude is the largest when $t=0$. From soliton theory, we know that the two solitary waves have elastic collision. Similarly, one can look at the breather solution shown in $(a)$ of Fig. 3 is a special bound state two-soliton solution formed by two solitary waves with the same wave velocity, wave width and amplitude, and has a periodic motion with respect to time $t$. The $(b)$ of Fig. 2 and Fig. 3 show the error dynamics of the difference between the exact solution and the predicted solution for the two-soliton solution and the breather solution, respectively. In Fig. 4, the corresponding three dimensional motion of the two-soliton solution and the breather solution are shown, respectively. It is evident that the breather solution is more symmetric than the general two-soliton solution.
\begin{figure}
\centering
\includegraphics[width=6.5cm,height=5cm]{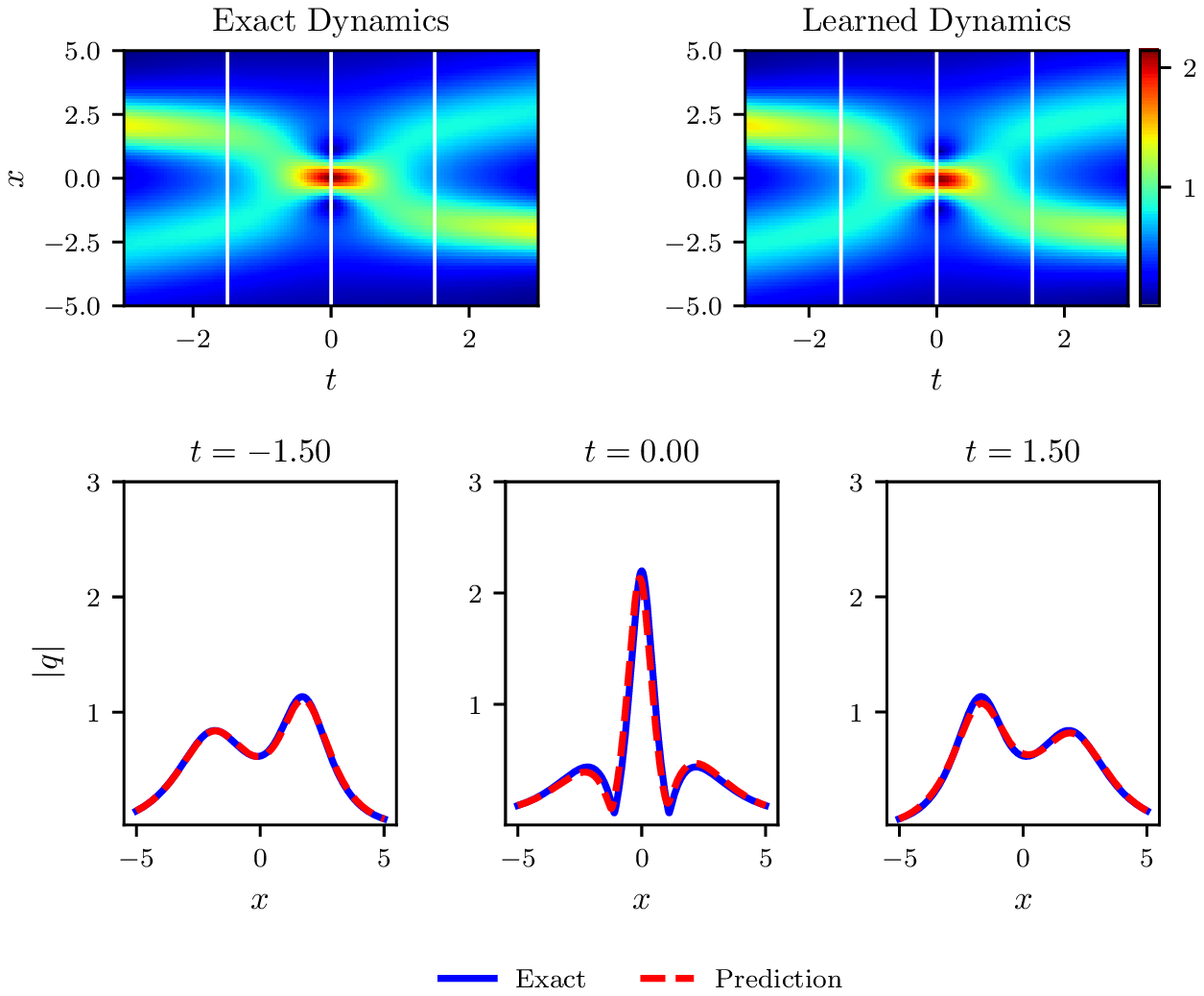}
$a$
\includegraphics[width=6.5cm,height=5cm]{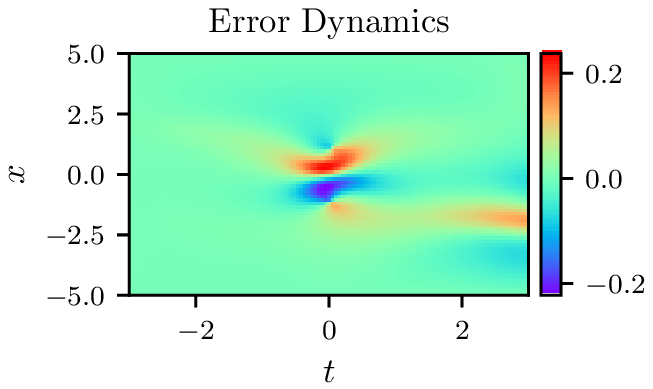}
$b$
\caption{The two-soliton solution $q(x,t)$: (a) The density diagrams and the profiles at different moments; (b) The error density diagram.}
\end{figure}

\begin{figure}
\centering
\includegraphics[width=6.5cm,height=5cm]{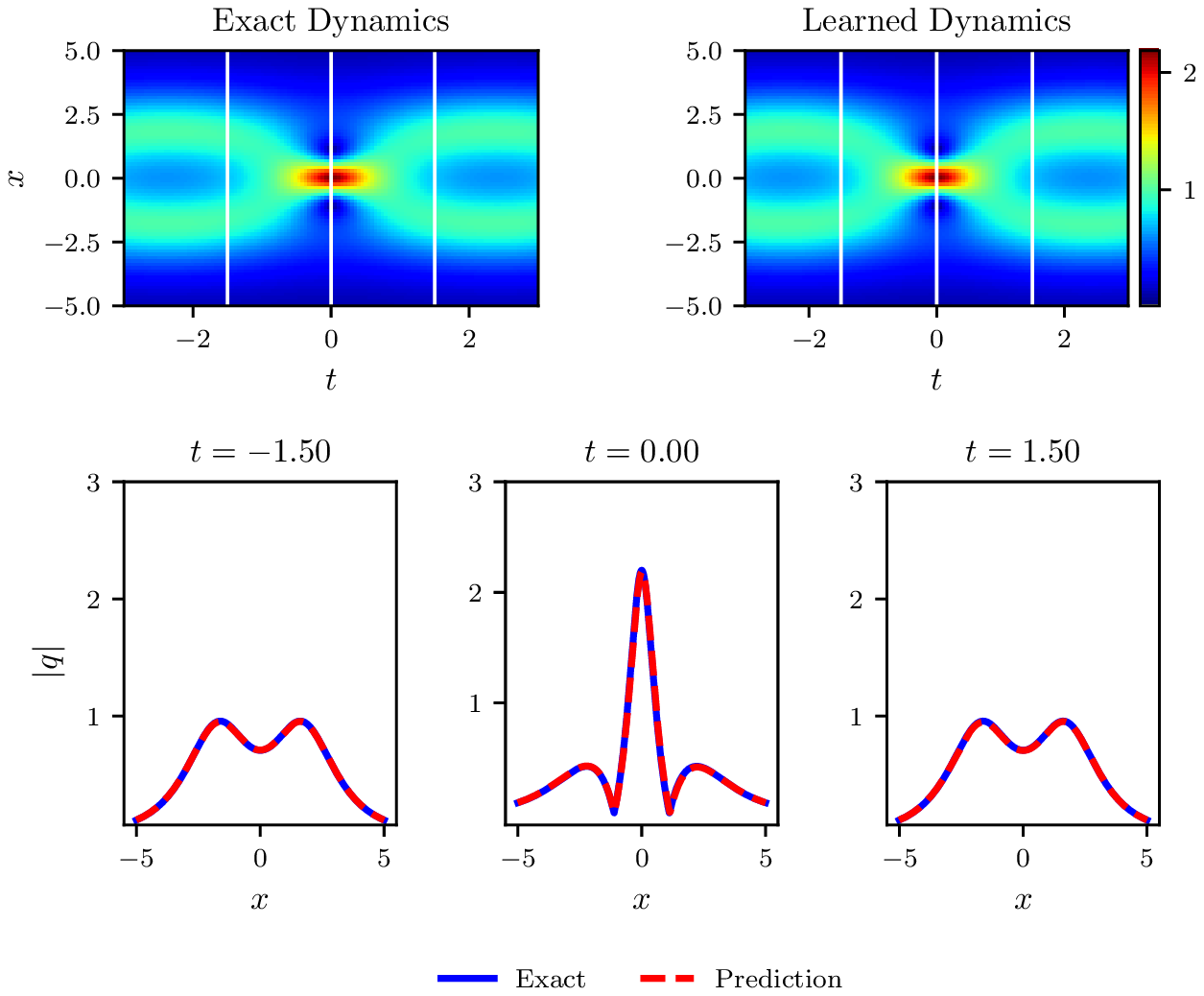}
$a$
\includegraphics[width=6.5cm,height=5cm]{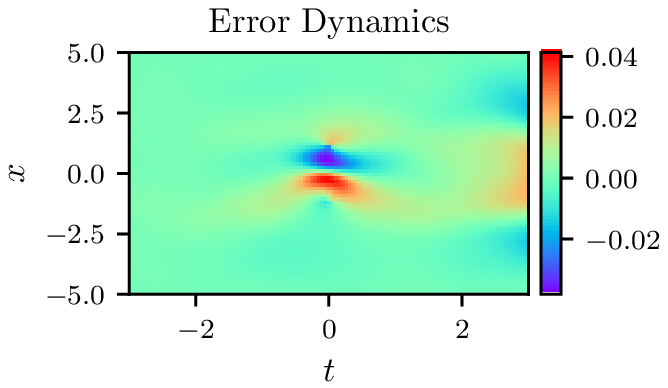}
$b$
\caption{The breather solution $q(x,t)$: (a) The density diagrams and the profiles at different moments; (b) The error density diagram.}
\end{figure}

\begin{figure}
\centering
\includegraphics[width=6.5cm,height=5cm]{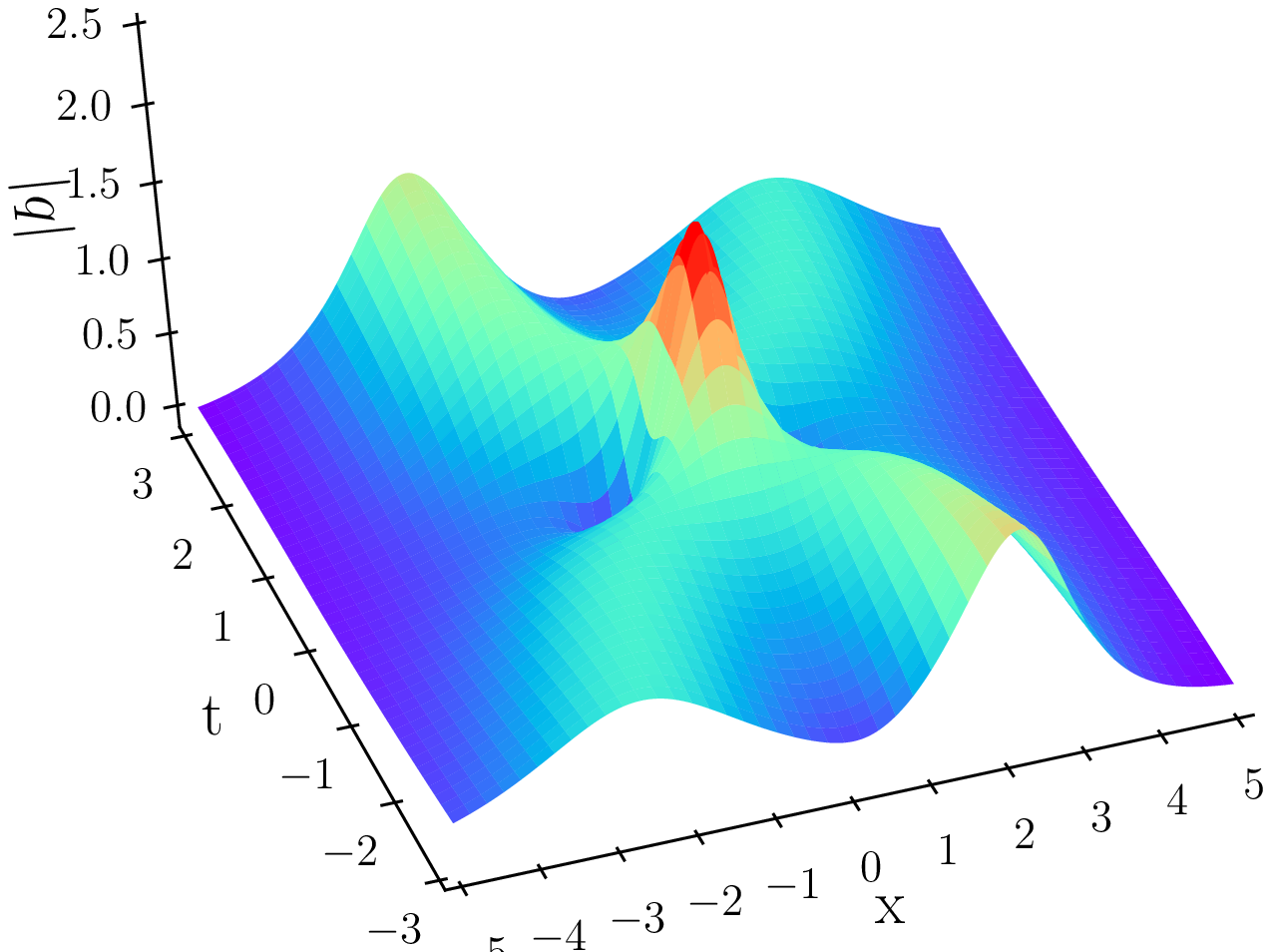}
$a$
\includegraphics[width=6.5cm,height=5cm]{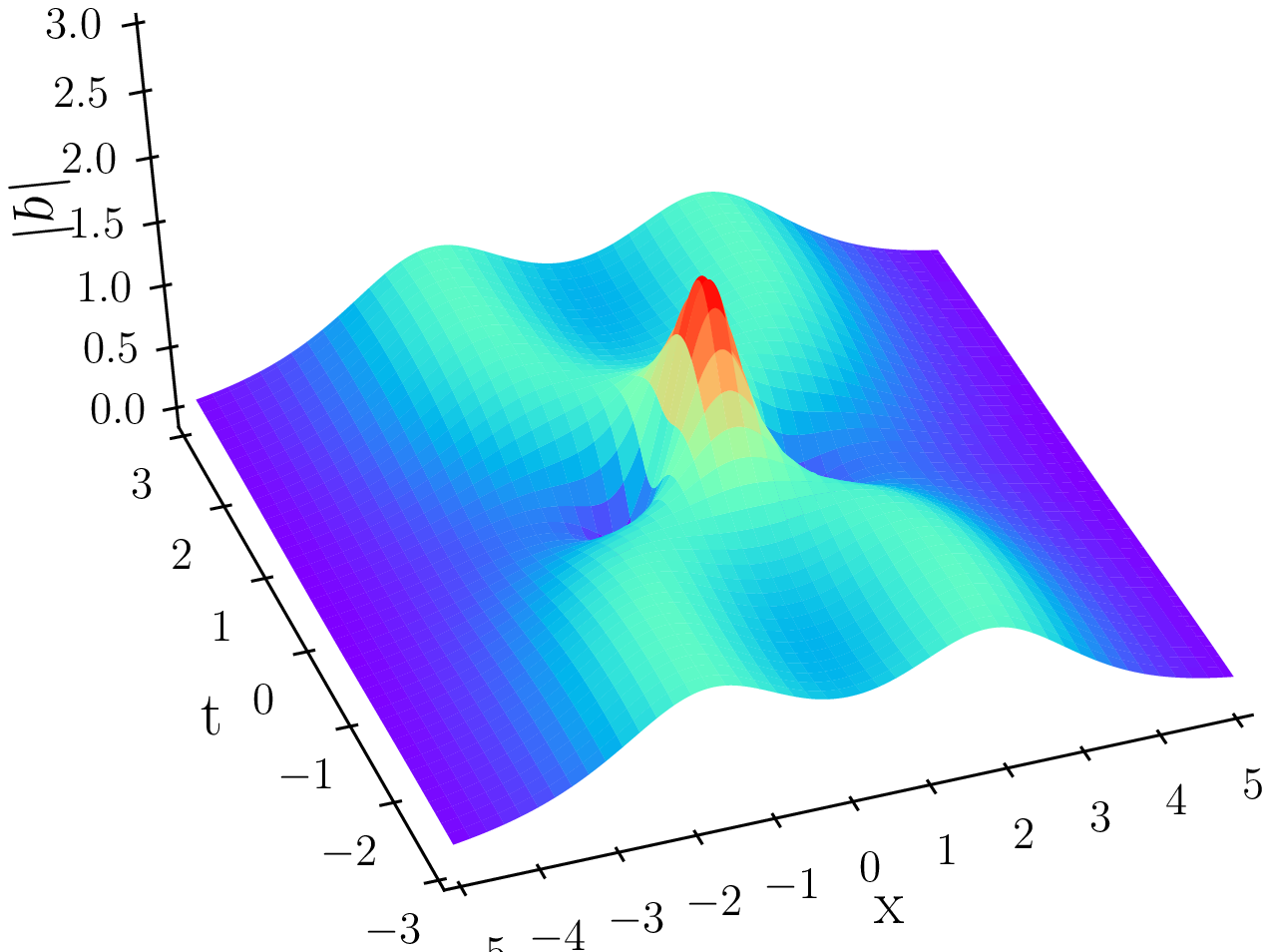}
$b$
\caption{The three dimensional motion of $q(x,t)$: (a) the two-soliton solution; (b) the breather solution.}
\end{figure}

For the numerical simulation of the three-soliton solution, we only need to take $N=3$ in Eqs. \eqref{E12}-\eqref{E15} to get the exact solution of the three-soliton solution, and then discretize the initial and boundary value data of the exact solution as our original dataset and train our network to simulate the corresponding three-soliton solution numerically. Similarly, N-soliton solutions can be learned by the same approach. Of course, the higher the order of soliton solution, the more complex the form of the solution, then the longer the resulting network training time takes.

\section{Rogue wave solutions of the nonlinear Schr\"odinger equation}

Recently, the research of rogue wave has been one of the hot topics in many areas such as optics, ocean dynamics, plasma, Bose-Einstein condensate and even finance \cite{ Solli2007,Chabchoub2011,Bludov2009,Moslem2011,Yan2011}. In addition to the peak amplitude more than twice of the background wave, rogue waves also have the characteristics of instability and unpredictability. Therefore, the study and application of rogue waves play a momentous role in real life, especially in avoiding the damage to ships caused by ocean rogue waves. As an one-dimensional integrable scalar equation, the nonlinear Schr\"odinger equation plays a key role in describing rogue waves. In 1983, Peregrine \cite{Peregrine1983} first gave a rational rogue waves to the nonlinear Schr\"odinger equation, whose generation principle is identified as the evolution of the breather waves when the period tends to infinity. At present, the researches on rogue wave of this equation through data-driven methods, such as machine learning, are relatively less. Marcucci et al.  \cite{Marcucci2020} have studied the computational machine in which nonlinear waves replace the internal layers of neural networks, discussed learning conditions, and demonstrated functional interpolation, datasets, and Boolean operations. When considering the solitons, rogue waves and shock waves of the nonlinear Schr\"odinger equation, highly nonlinear and even discontinuous regions play a leading role in the network training and solution calculation. In this section, we construct the rogue wave solutions of the nonlinear Schr\"odinger equation by the neural network with underlying physical constraints. Here, we consider the another form of focusing nonlinear Schr\"odinger equation along with Dirichlet boundary conditions given by
\begin{equation}\label{E27}
\begin{split}
\begin{cases}
iq_t+\frac{1}{2}q_{xx}+|q|^2q=0,x\in[x_0,x_1],t\in[t_0,t_1],\\
q(x,t_0)=q_0(x),\\
q(x_0,t)=q(x_1,t),\\
\end{cases}
\end{split}
\end{equation}
where $q_0(x)$ is an arbitrary complex-valued function of space variable $x$, here $x_0,x_1$ represent the lower and upper boundaries of $x$ respectively, and $t_0,t_1$ represent the initial and terminal time instants of $t$ respectively. In addition, this equation corresponds to Eq. \eqref{E1} with $\alpha=\frac{1}{2}$ and $\beta=1$. The rogue wave solutions of Eq. \eqref{E27} can be obtained by lots of different tools \cite{Akhmediev2009}. Therefore, we can get respectively the one-order rogue wave and the two-order rogue wave of Eq. \eqref{E27} as follows
\begin{equation}\label{E28}
q(x,t)=\left[1-\frac{4(1+2it)}{4t^2+4x^2+1}\right]e^{it},
\end{equation}
and
\begin{equation}\label{E29}
q(x,t)=\left(1+\frac{G+itH}{D}\right)e^{it},
\end{equation}
where
\begin{equation}\nonumber
\begin{split}
&G=\frac38-3x^2-2x^4-9t^2-10t^4-12t^2x^2,\\
&H=\frac{15}{4}+6x^2-4x^4-2t^2-4t^4-8t^2x^2,\\
&D=\frac{3}{32}+\frac98x^2+\frac12x^4+\frac23x^6+\frac{33}{8}t^2+\frac92t^4+\frac23t^6-3t^2x^2+2t^2x^4+2t^4x^2.
\end{split}
\end{equation}

In the two following parts, we will construct the training dataset to reconstruct our predicted solutions based on the above two rogue wave solutions by constructing a neural network with 9 hidden layers and 40 neurons per hidden layer.

\subsection{One-order rogue wave}
\quad

In this subsection, we will numerically uncover the one-order rogue wave of the nonlinear Schr\"odinger equation using the neural network method above. Now, we take $[x_0,x_1]$ and $[t_0,t_1]$ in Eq. \eqref{E27} as $[-2.0,2.0]$ and $[-1.5,1.5]$, respectively. The corresponding initial condition is obtained from \eqref{E28}, we have
\begin{equation}\label{E30}
q_0(x)=\left(1+\frac{-4+12i}{4x^2+10}\right)e^{-1.5i}.
\end{equation}

Next, we obtain the initial and boundary value dataset by the same data discretization method in Section 3.1, and then we can simulate precisely the one-order rogue wave solution by feeding the data into the network. By randomly subsampling $N_q=100$ from the original dataset and selecting $N_f=10000$ configuration points which are generated by LHS, a training dataset composed of initial-boundary data and collocation points is generated. After training, the neural network model achieves a relative $\mathbb{L}_2$ error of 7.845201$\mathrm{e-}$03 in about $871$ seconds, and the number of iterations is 9584.

Our experiment results are summarized in Fig. 5, and we simulate the solution $q(x,t)$ and then obtain the density diagrams, profiles at different instants, error dynamics diagrams, three dimensional motion and loss curve figure of the one-order rogue wave. Specifically, the magnitude of the predicted spatio-temporal solution $|q(x,t)|$ is shown in the top panel of (a) of Fig. 5. It can be simply seen that the amplitude of the rogue wave solution changes greatly in a very short time from the bottom panel of (a) in Fig. 5. Meanwhile, we present a comparison between the exact and the predicted solution at different time instants $t=-0.75, 0, 0.75$. (b) of Fig. 5 reveals the relative $\mathbb{L}_2$ error becomes larger as the time increases. From (d) of Fig. 5, we can observe that when the number of iterations is more than 2000, there are some obvious fluctuations which we could call "burr" in the training, it does not exist during the training process about the one-soliton solution of the nonlinear Schr\"odinger equation. With only a handful of initial-boundary data, one can accurately capture the intricate nonlinear dynamical behavior of the integrable Schr\"odinger equation by this method.
\begin{figure}
\centering
\includegraphics[width=6.5cm,height=5cm]{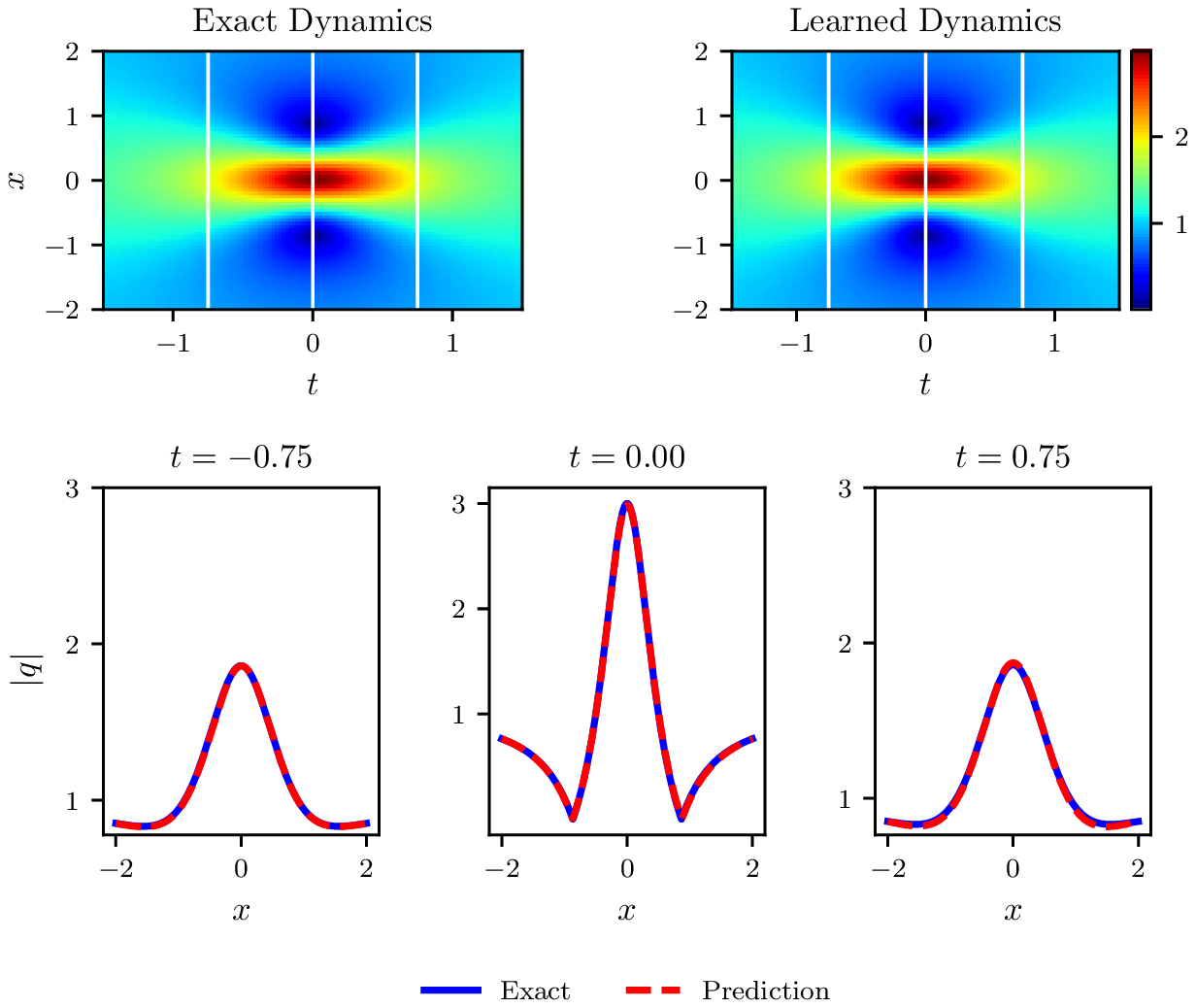}
$a$
\includegraphics[width=6.5cm,height=5cm]{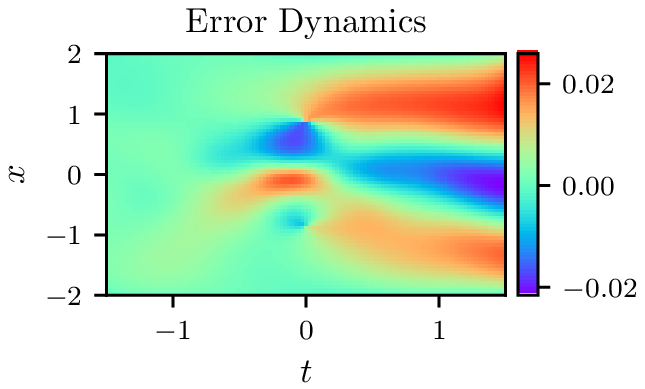}
$b$\\
\includegraphics[width=6.5cm,height=5cm]{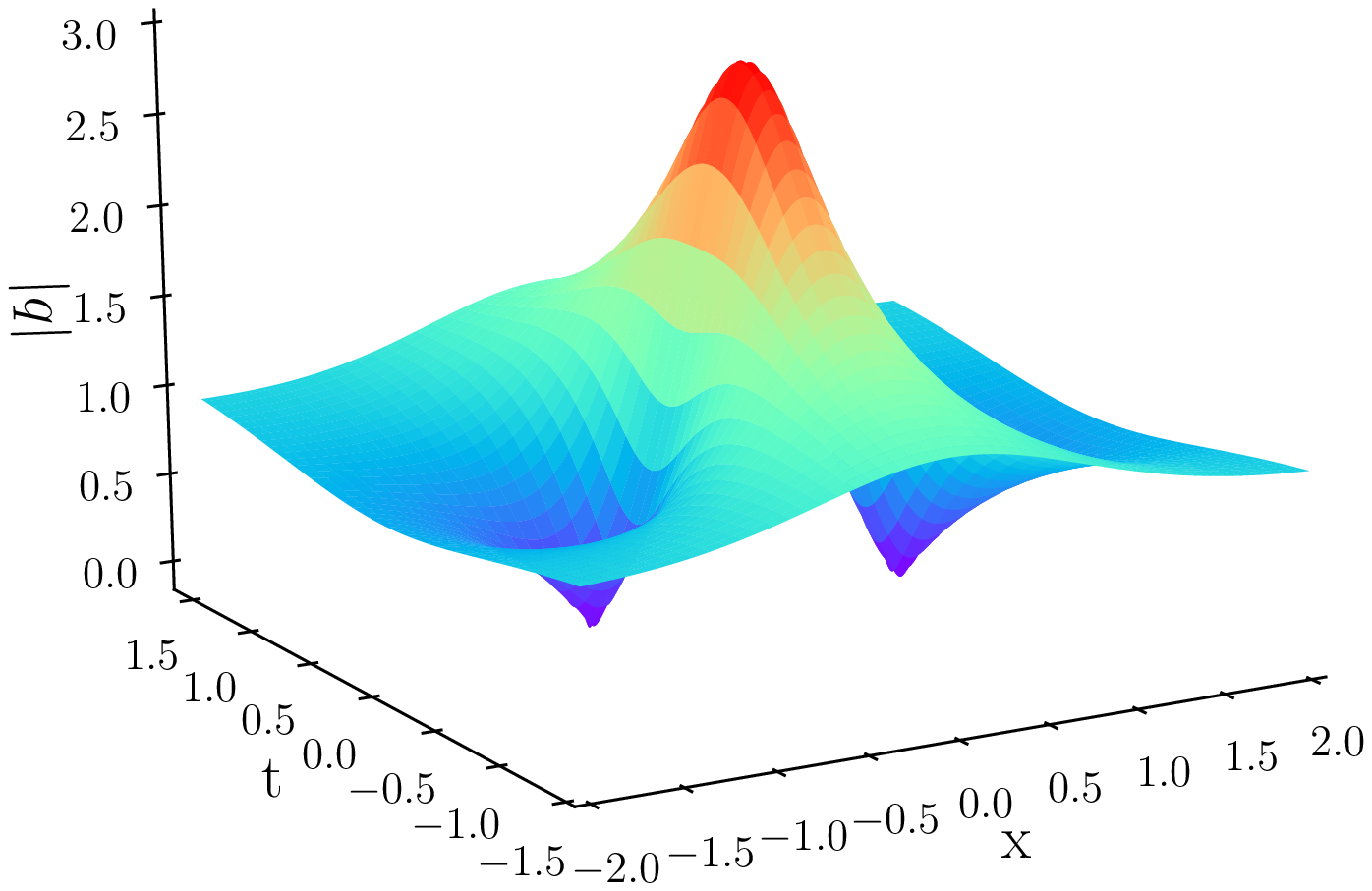}
$c$
\includegraphics[width=6.5cm,height=5cm]{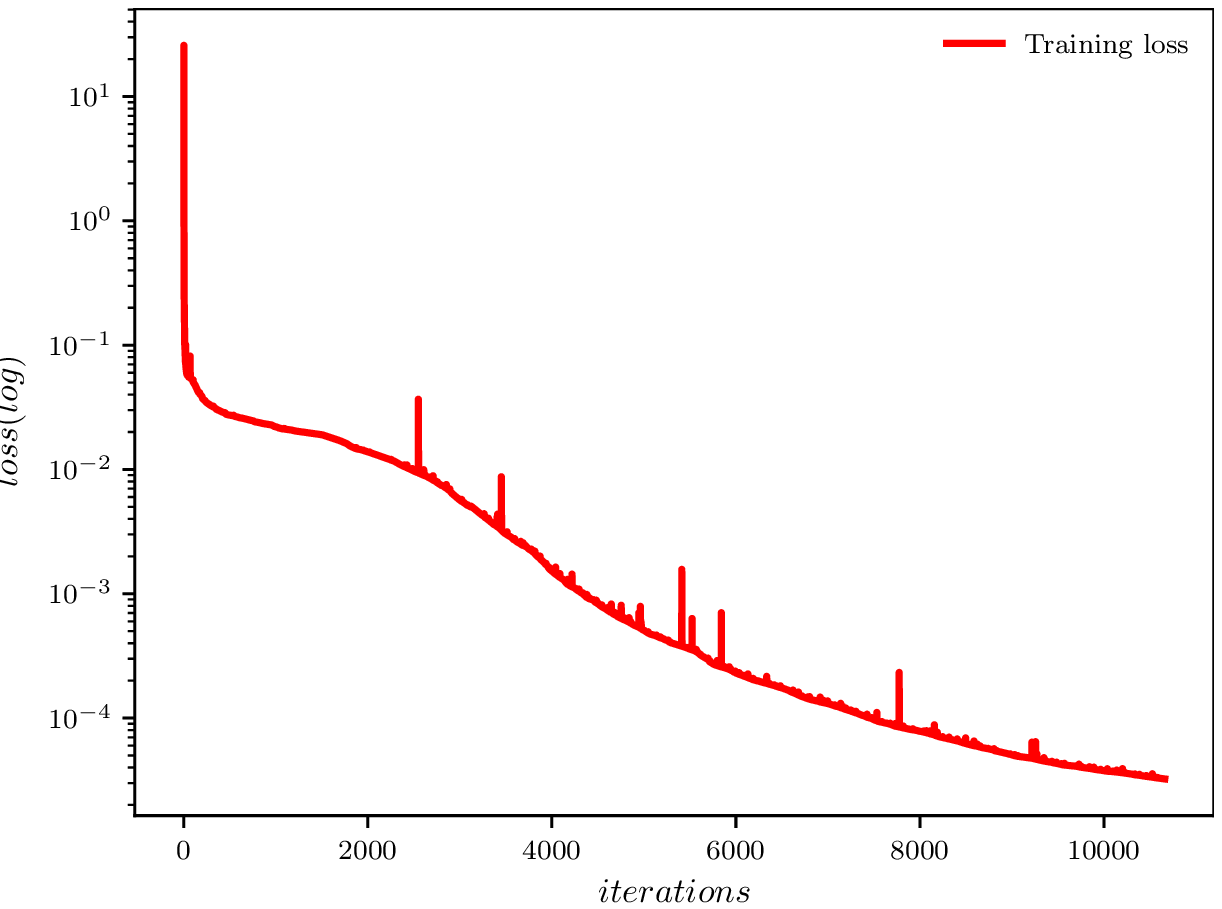}
$d$
\caption{The one-order rogue wave solution $q(x,t)$: (a) The density diagram and profiles at three different instants; (b) The error density diagram; (c) The three-dimensional motion; (d) The loss curve.}
\end{figure}

In addition, based on the same initial and boundary values of the one-order rogue waves in the case of $N_q=100$ and $N_f=10000$, we employ the control variable method often used in applied sciences to study the effects of different numbers of network layers and neurons per hidden layer on the one-order rogue wave dynamics of nonlinear Schr\"odinger equation. The relative $\mathbb{L}_2$ errors of different network layers and different neurons per hidden layer are given in Table 1. From the data in Table 1, we can see that when the number of network layers is fixed, the more the number of single-layer neurons, the smaller the relative error becomes. Due to the influence of randomness caused by some factors, there are some cases that do not conform with the above conclusion. However, when the number of single-layer neurons is fixed, the influence of the number of network layers on the relative error is not obvious. To sum up, we can draw the conclusion that the network layers and the single-layer neurons jointly determine the relative $\mathbb{L}_2$ error to some extent. In the case of the same training dataset, Table 2 shows the relative $\mathbb{L}_2$ error with 9 network layers and 40 neurons per hidden layer when taking different numbers of subsampling points $N_q$ in the initial-boundary data and collocation points $N_f$. From the Table 2, we can see that the influence of $N_q$ on the relative $\mathbb{L}_2$ error of the network is not obvious, which also indicates the network model with physical constraints can uncover accurate predicted solutions with smaller initial-boundary data and relatively many sampled collocation points.

\begin{table}[htbp]
  \caption{One-order rogue wave of the nonlinear Schr$\mathrm{\ddot{o}}$dinger equation: Relative final prediction error estimations in the $\mathbb{L}_2$ norm for different numbers of network layers and neurons per hidden layer.}
  \centering
  \begin{tabular}{p{3.38cm}|p{2cm}p{2cm}p{2cm}p{2cm}p{2cm}}
  \toprule
  \textbf{\diagbox{Layers}{Neurons}} &\textbf{\quad\quad20} &\textbf{\quad\quad30} &\textbf{\quad\quad40} &\textbf{\quad\quad50} &\textbf{\quad\quad60}\\
  \midrule
  \textbf{5}   &2.765905e-03&2.903368e-04&4.961406e-04&5.232502e-04&9.323978e-04\\
  \textbf{7}   &2.699082e-03&1.328768e-03&4.030658e-04&3.633812e-04&1.448091e-03\\
  \textbf{9}   &2.732954e-03&2.618465e-03&7.845201e-03&6.880915e-04&6.797486e-04\\
  \textbf{11}  &4.641999e-03&1.779715e-03&1.440061e-03&9.148106e-04 &1.581767e-03\\
  \bottomrule
  \end{tabular}
\end{table}

\begin{table}[htbp]
  \caption{One-order rogue wave of the nonlinear Schr$\mathrm{\ddot{o}}$dinger equation: Relative final prediction error measurements in the $\mathbb{L}_2$ norm for different numbers of $N_q$ and $N_f$.}
  \centering
  \begin{tabular}{p{1.35cm}|p{2cm}p{2cm}p{2cm}p{2cm}p{2cm}}
  \toprule
  \textbf{\diagbox{$N_q$}{$N_f$}} &\textbf{\quad6000} &\textbf{\quad8000} &\textbf{\quad10000} &\textbf{\quad12000} &\textbf{\quad14000}\\
  \midrule
  \textbf{80}    &1.473695e-02&9.200569e-03&7.967294e-03&6.034213e-03&3.575678e-03\\
  \textbf{100}   &1.176106e-02&3.082057e-03&7.845201e-03&5.495886e-03&1.332274e-02\\
  \textbf{120}   &1.525780e-02&1.265775e-02&4.175621e-02&2.402183e-03&6.568740e-03\\
  \bottomrule
  \end{tabular}
\end{table}

\subsection{Two-order rogue wave}
\quad

In the next example, we consider the two-order rogue wave of the nonlinear Schr\"odinger equation, and properly take $[x_0,x_1]$ and $[t_0,t_1]$ in Eq. \eqref{E27} as $[-2.0,2.0]$ and $[-0.5,0.5]$. Here we consider the corresponding initial condition from \eqref{E29} as follows
\begin{equation}\label{E31}
q_0(x)=\left[1+\frac{-70.5-30x^2-2x^4-1.5i(-21-12x^2-4x^4)}{39.75+4.5x^2+5x^4+\frac23x^6}\right]e^{-1.5i}.
\end{equation}

We use the same data discretization method in Section 3.1 to collect the initial and boundary data. In the network architecture, initial and boundary training dataset of $N_q=100$ are randomly subsampled from the original initial-boundary data. In addition, configuration points of $N_f=10000$ are sampled by LHS. Finally, the hidden two-order rogue wave solution of nonlinear Schr\"odinger equation is approximated fairly accurately by constraining the loss function with underlying physical laws. The neural network model achieves a relative $\mathbb{L}_2$ error of 1.665401$\mathrm{e-}$02 in about $1090$ seconds, and the number of iterations is 11450.

The detailed illustration is shown in Fig. 6. The top panel of (a) of Fig. 6 gives the density map of hidden solution $q(x,t)$, and when combing (b) of Fig. 6 with the bottom panel of (a) in Fig. 6, we can see that the relative error is relatively large at t = 0.25. From (d) of Fig. 6, in contrast with the one-order rogue wave solution, the fluctuation (burr phenomenon) of the loss function is obvious when the number of iterations is less than 3000.

\begin{figure}
\centering
\includegraphics[width=6.5cm,height=5cm]{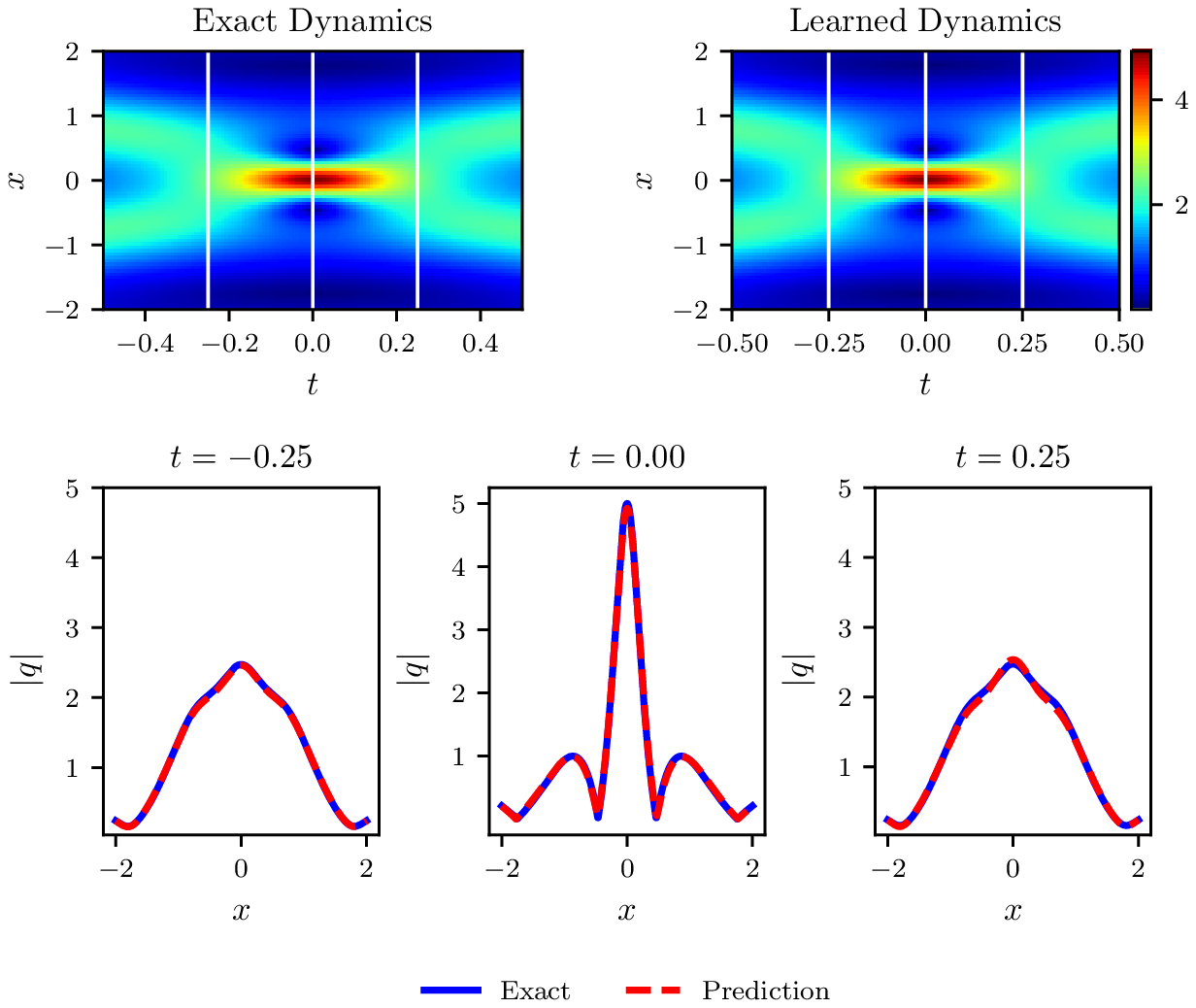}
$a$
\includegraphics[width=6.5cm,height=5cm]{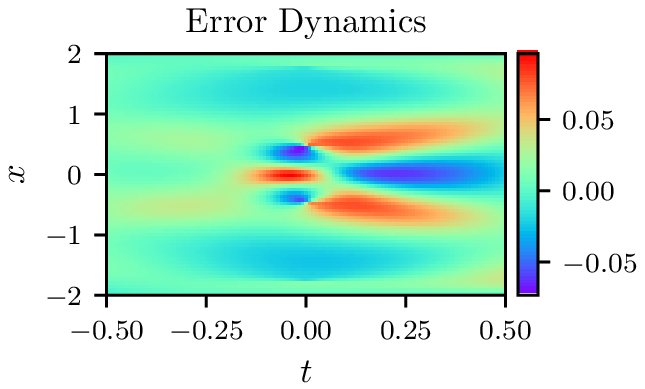}
$b$\\
\includegraphics[width=6.5cm,height=5cm]{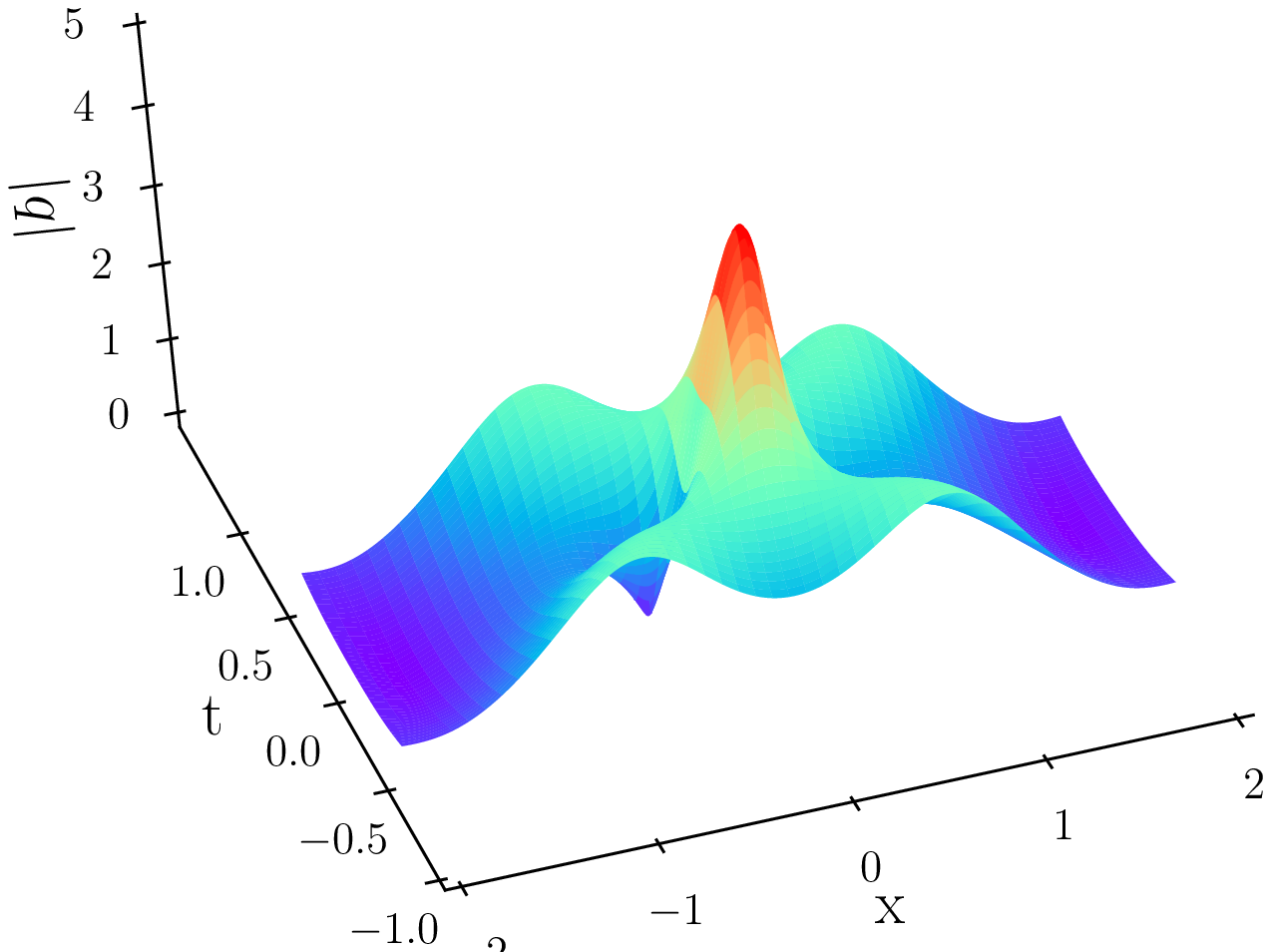}
$c$
\includegraphics[width=6.5cm,height=5cm]{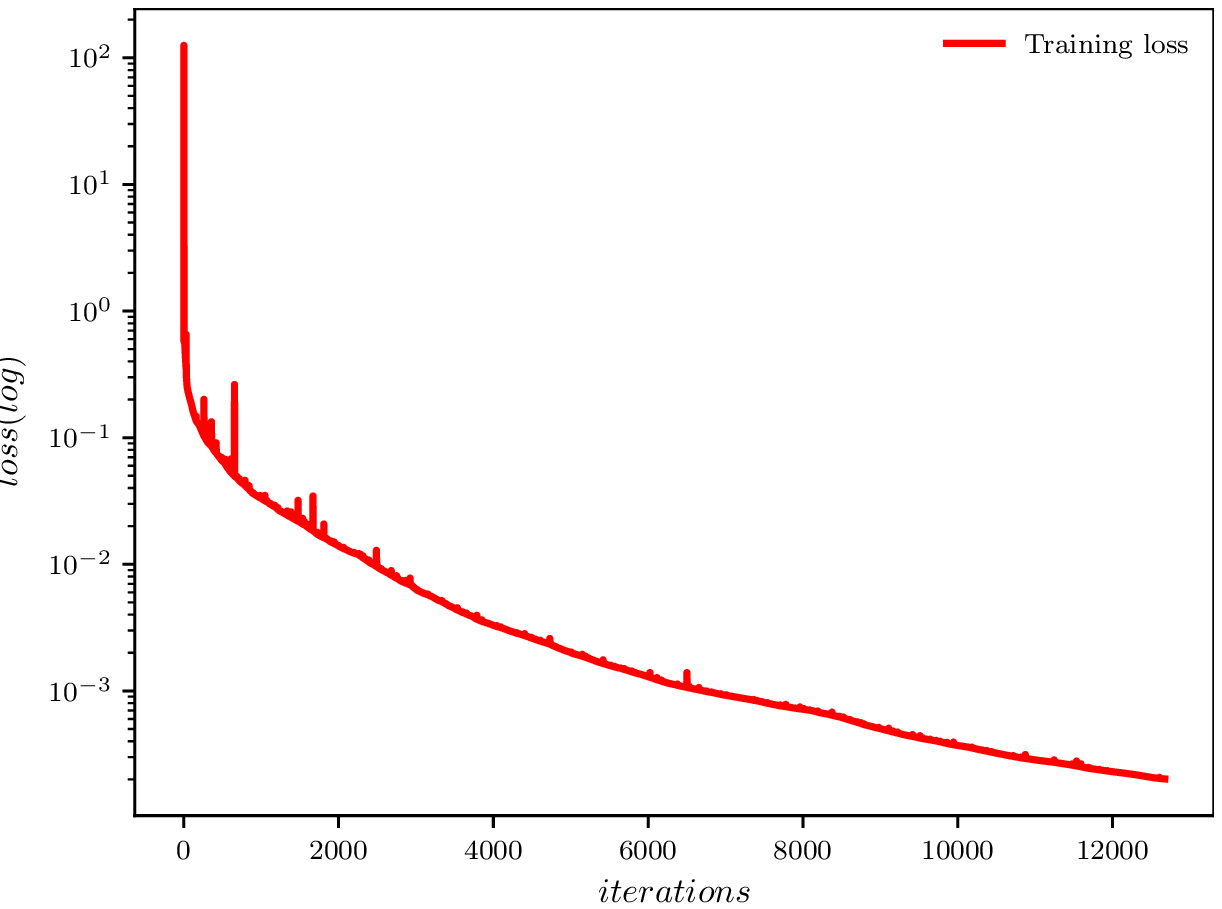}
$d$
\caption{The two-order rogue wave solution $q(x,t)$: (a) The density diagrams and the snapshots at three different instants; (b) The error density diagram; (c) The three-dimensional motion; (d) The loss curve figure.}
\end{figure}

\section{Conclusion}

In this paper, we introduce a physically-constrained deep learning method based on PINN to solve the classical integrable nonlinear Schr\"odinger equation. Compared with traditional numerical methods, it has no mesh size limits. Moreover, due to the physical constraints, the network is trained with just few data and has a better physical interpretability. This method showcases a series of results of various problems in the interdisciplinary field of applied mathematics and computational science which opens a new path for using deep learning to simulate unknown solutions and correspondingly discover the parametric equations in scientific computing.

Specifically, we apply the data-driven algorithm to deduce the soliton solutions, breather solution and rogue wave solutions to the nonlinear Schr\"odinger equation. We outline how different types of solutions (such as general soliton solutions, breather solution and rogue wave solutions) are generated due to different choices of initial and boundary value data. Remarkably, these results show that the deep learning method with physical constraints can exactly recover different dynamical behaviors of this integrable equation. Furthermore, the sizes of space-time variable $x$ and $t$ interval are selected by the dynamical behaviors of these solutions. For the breathers, in particular, the wider the interval of time variable $t$, the better we can see the dynamical behavior in this case. However, with a wider range of time interval $t$, the training effect is not very good. So more complex boundary conditions, such as Neumann boundary conditions, Robin boundary conditions or other mixed boundary conditions, may be considered. Similarly, for the integrable complex modified Korteweg-de Vries (mKdV) equation, the Dirichlet boundary conditions can not recover the ideal rogue wave solutions.

The influence of noise on our neural network model is not introduced in this paper. This kind of physical factors in real life should be considered to show the network's robustness. Compared with static LHS sampling with even mesh sizes, more adaptive sampling techniques should be considered in some special problems, for example, discontinuous fluid flows such as shock wave. In addition, more general nonlinear Schr\"odinger equation, such as the derivative Schr\"odinger equation, is not investigated in this work. These new problems and improvements will be considered in the future research.

\end{document}